\documentclass[preprint,showpacs,aps,pra,amsmath,amssymb,floatfix]{revtex4}
\usepackage{graphicx}
\usepackage{dcolumn}
\usepackage{bm}
\usepackage{setspace}
\usepackage{amsmath}
\usepackage{amsfonts}
\usepackage{amssymb}
\providecommand{\U}[1]{\protect\rule{.1in}{.1in}}

\begin{document}

\title{Power counting of various Dirac covariants in hadronic Bethe-Salpeter wave
functions for decay constant calculations of pseudoscalar mesons}

\author{Shashank Bhatnagar\footnote{Regular Associate, ICTP, e-mail: {\tt
shashank\_bhatnagar@yahoo.com}}}

\affiliation{Department of Physics, Addis Ababa University,
P.O.Box 1148/1110,
Addis Ababa, Ethiopia and\\
The Abdus Salam ICTP, Strada Costiera 11, 34100, Trieste, Italy}

\author{Shi-Yuan Li  \footnote{e-mail: {\tt lishy@sdu.edu.cn}}}

\affiliation{School of Physics, Shandong University, Jinan,
250100, P. R. China}

\author{Jorge Mahecha\footnote{Senior Associate, ICTP, e-mail: {\tt
mahecha@fisica.udea.edu.co}}}

\affiliation{Instituto de F\'{\i}sica, Universidad de Antioquia,
Calle 67
No 53-108, AA 1226, Medellin, Colombia and\\
The Abdus Salam ICTP, Strada Costiera 11, 34100, Trieste, Italy}

\begin{abstract}
We have employed the framework of Bethe-Salpeter equation under
covariant instantaneous ansatz to calculate leptonic decay
constants of unequal mass pseudoscalar mesons like $\pi^\pm$, $K$,
$D$, $D_s$ and $B$ and radiative decay constants of neutral
pseudoscalar mesons like $\pi^0$ and $\eta _c$ in two photons. In
the Dirac structure of hadronic Bethe-Salpeter wave function, the
covariants are incorporated from their complete set in accordance
with a recently proposed power counting rule. The decay constants
are calculated with the incorporation of both Leading order and
Next-to-leading order  Dirac covariants. The results  validate the
power counting rule which  provides a practical means of
incorporating Dirac covariants in the Bethe-Salpeter wave function
for a hadron.
\end{abstract}

\maketitle

\section{Introduction}

Quantum Chromodynamics (QCD) is the theory to describe strong interactions.
However, the large gauge coupling at low energies (long distances) destroys
the perturbative expansion. As a result, many non-perturbative approaches have
been proposed to deal with this long distance properties of QCD, such as QCD
sum rules, Lattice QCD, dynamical-equation-based approaches like
Schwinger-Dyson equation and Bethe-Salpeter equation (BSE), and potential
models.
%
%
Since the task of calculating hadron structures from QCD itself is very
difficult, as can be seen from various Lattice QCD approaches, one generally
relies on specific models to gain some understanding of QCD at low energies.
BSE is a conventional approach in dealing with relativistic bound state
problems. From the solutions we can obtain useful information about the inner
structure of hadrons, which is also crucial in treating high energy hadronic
scatterings. The BSE framework is firmly rooted in field theory, and provides
a realistic description for analyzing hadrons as composite objects. Despite
its drawback of having to input model-dependent kernel, these studies have
become an interesting topic in recent years, since calculations have shown
that BSE framework using phenomenological potentials can give satisfactory
results on more and more data being accumulated.

In this paper we study leptonic decays of pseudoscalar mesons (P-mesons) such
as $\pi$, $K$, $D$, $D_{S}$ and $B$, which proceed through the coupling of
quark-antiquark loop to the axial vector current and also the two-photon
decays of neutral pseudoscalar mesons such as $\pi^{0}$and $\eta_{c}$ which
proceed through the famous quark-triangle diagrams. We employ QCD motivated
BSE under Covariant Instantaneous Ansatz (CIA) in this paper \cite{a,
bhatnagar06}. CIA is a Lorentz-invariant generalization of Instantaneous
Ansatz. For a $\mathrm{q}\overline{\mathrm{q}}$ system, the CIA formulation
ensures an exact interconnection between 3D and 4D forms of BSE
\cite{bhatnagar06, bhatnagar09}. The 3D form of BSE serves for making contact
with the mass spectrum, whereas the 4D form provides the Hadron-quark vertex
function for evaluation of various hadronic transition amplitudes through
quark loop diagrams. In these studies one of the main ingredients is the Dirac
structure of the Bethe-Salpeter wave function (BSW). The copious Dirac
structure of BSW was already studiedd by Llewllyn Smith \cite{smith69} much
earlier. Recent studies \cite{cvetic04,alkofer02} have revealed that various
covariant structures in BSWs of various hadrons is necessary to obtain
quantitatively accurate observables. It has been further noticed that all
covariants do not contribute equally for calculation of meson observables. So
it is interesting to investigate how to arrange these covariants. In a recent
work \cite{bhatnagar06}, we developed a power counting rule for incorporating
various Dirac structures in BSW, order-by-order in powers of inverse of meson
mass. We have outlined the Dirac covaiants and expanded the coefficients to
the leading order (LO), and calculated the leptonic decay constants of vector
mesons ($\rho$, $\omega$, $\phi$, $\psi$) \cite{bhatnagar06} as well as
pseudoscalar mesons ($\pi$, $K$, $D$, $D_{S}$ and $B$) \cite{bhatnagar09} at
this order.
%
%
The results agree with data well.

However, common to all the perturbative theories, it is better to calculate
the next order(s) to the leading one and make sure it is (they are) really
smaller w.r.t. the LO, before claiming the validation of the perturbation. At
the same time, as more and preciser data accummulated, it is useful to arrange
more available parameters inherent in our framework to accommodate better fits
to gain more precise information of the structure of hadron. So the study of
next-to-leading order (NLO) is natural and essential. For all the mesons, the
pseudoscalar is the simplest in Dirac structure.
%
As the first step, we collect the data of leptonic decay constants $f_{P}$'s
for pseudoscalar mesons ($\pi$, $K$, $D$, $D_{S}$ and $B$), to fit three
parameters $B_{i\text{ }}^{\prime}s$ in our framework at NLO. We found: a) NLO
works better than LO. b) NLO corrections are smaller than those of LO ($\pi$
is exceptional for its small mass, to be discussed later in this paper). Then
with the fitted parameters we calculate the radiative decay constants $F_{P}$
of neutral pseudoscalar mesons, $\pi^{0}$ and $\eta_{c}$ at NLO. We also found
satisfying agreement with data, and fair improvement w.r.t. LO. Thus the fact
that three parameters can give a good fit not only for 5 different cases of
$f_{P}$, but also giving satisfactory results for two cases of $F_{P}$,
demonstrates the validity and robustness of this framework.
%
%
These results indicate that our power counting scheme \cite{bhatnagar06}
provides a practical means of incorporating various Dirac structures from
their complete set into the BS wave function.

In what follows, we give a detailed discussion of the fit and calculation at
NLO, after a brief review of our framework.
%
The paper is organized as follows: In section \ref{sec:2}, we discuss the
structure of BS wave function for P-mesons in BSE under CIA using the power
counting rule. In section \ref{sec:3}, we introduce the fitting to $f_{P}$ for
pseudoscalar mesons. The radiative decay constants $F_{P}$ for $\pi^{0}$and
$\eta_{c}$ mesons are calculated in section 4, while we conclude with
Discussion in section \ref{sec:5}.

%

\section{The BSW under CIA}

\label{sec:2}

\subsection{BSE under CIA}

\label{subsec:CIA}

We first outline the BSE framework under CIA. We have employed for the case of
scalar quarks for simplicity. For a $\mathrm{q}\overline{\mathrm{q}}$ system
with an effective kernel $K$ and 4D wave function $\Phi(P,q)$, the 4D BSE
takes the form,
\begin{equation}
i(2\pi)^{4}\Delta_{1}\Delta_{2}\Phi(P,q)=\int d^{4}qK(q,q^{\prime}%
)\Phi(P,q^{\prime}), \label{eq:2.1}%
\end{equation}
where $\Delta_{1,2}=m_{1,2}^{2}+p_{1,2}^{2}$are the inverse propagators of two
scalar quarks, and $m_{1,2}$ are (effective) constituent masses of quarks. The
4-momenta of the quark and anti-quark, $p_{1,2}$, are related to the internal
4-momentum $q_{\mu}$ and total momentum $P_{\mu}$ of hadron of mass $M$ as
\begin{equation}
p_{1,2}{}_{\mu}=\widehat{m}_{1,2}P_{\mu}\pm q_{\mu}, \label{eq:2.3}%
\end{equation}
where $\widehat{m}_{1,2}=[1\pm(m_{1}^{2}-m_{2}^{2})/M^{2}]/2$ are the
Wightman-Garding (WG) definitions of masses of individual quarks. Now it is
convenient to express the internal momentum of the hadron $q$ as the sum of
two parts, the transverse component, $\hat{q}_{\mu}=q_{\mu}-\frac{q.P}{P^{2}}$
which is orthogonal to total hadron momentum $P$ (ie. $\widehat{q}.P=0$
regardless of whether the individual quarks are on-shell or off-shell), and
the longitudinal component, $\sigma P_{\mu}=(q\cdot P/P^{2})P_{\mu}$, which is
parallel to P. We now use an Ansatz on the BS kernel $K$ in Eq. (\ref{eq:2.1})
which is assumed to depend on the 3D variables $\hat{q}_{\mu}$, $\hat{q}_{\mu
}^{\prime}$ \cite{mitra92} i.e.
\begin{equation}
K(q,q^{\prime})=K(\hat{q},\hat{q}^{\prime}), \label{eq:2.4}%
\end{equation}
A similar form of the BS kernel was also earlier suggested in ref.
\cite{resag94}). Hence, the longitudinal component, $\sigma P_{\mu}$ of
$q_{\mu}$, does not appear in the form $K(\hat{q},\hat{q}^{\prime})$ of the
kernel. For reducing Eq.(\ref{eq:2.1}) to the 3D form, we define a 3D wave
function $\phi(\hat{q})$ as
\begin{equation}
\phi(\hat{q})=\int\limits_{-\infty}^{+\infty}{Md\sigma\Phi(P,q)}.
\label{eq:2.7}%
\end{equation}
Substituting Eq. (\ref{eq:2.7}) in Eq. (\ref{eq:2.1}), with definition of
kernel in Eq. (\ref{eq:2.4}), we get a covariant version of Salpeter
equation,
\begin{equation}
(2\pi)^{3}D(\hat{q})\phi(\hat{q})=\int d^{3}\bm{\hat{q}'}K(\hat{q},\hat
{q}^{\prime})\phi(\hat{q}^{\prime}), \label{eq:2.8}%
\end{equation}
where $D(\hat{q})$ is the 3D denominator function defined by
\begin{equation}
\frac{1}{D(\hat{q})}=\frac{1}{2\pi i}\int\limits_{-\infty}^{+\infty}%
\frac{Md\sigma}{\Delta_{1}\Delta_{2}}, \label{eq:2.9}%
\end{equation}
whose value can be easily worked out by contour integration by noting
positions of poles in the complex $\sigma$-plane (shown in detail in
\cite{bhatnagar05}) as,
\begin{equation}
D(\hat{q})=\frac{(\omega_{1}+\omega_{2})^{2}-M^{2}}{\displaystyle\frac
{1}{2\omega_{1}}+\frac{1}{2\omega_{2}}},\quad\omega_{1,2}^{2}=m_{1,2}^{2}%
+\hat{q}^{2}. \label{eq:2.10}%
\end{equation}
We can see that RHS of Eq. (\ref{eq:2.8}) is identical to RHS of Eq.
(\ref{eq:2.1}) by virtue of Equations (\ref{eq:2.4}) and (\ref{eq:2.7}). We
thus have an exact interconnection between 3D wave function $\phi(\hat{q})$
and 4D wave function $\Phi(P,q)$:
\begin{equation}
\Delta_{1}\Delta_{2}\Phi(P,q)=\frac{D(\hat{q})\phi(\hat{q})}{2\pi i}%
\equiv\Gamma(\hat{q}). \label{eq:2.11}%
\end{equation}
We also get the $H{q\bar{q}}$ vertex function $\Gamma(\hat{q})$ under CIA for
case of scalar quarks. Further in the process, an exact interconnection
between 3D and 4D BSE \cite{mitra92} is thus brought out where the 3D form
serves for making contact with the mass spectrum of hadrons, whereas the 4D
form provides the vertex $H{q\bar{q}}$ function $\Gamma(\hat{q})$ which
satisfies a 4D BSE with a natural off-shell extension over the entire 4D space
(due to the positive definiteness of the quantity $\hat{q}^{2}=q^{2}-(q\cdot
P)^{2}/P^{2}$ throughout the entire 4D space) and thus provides a fully
Lorentz-invariant basis for evaluation of various transition amplitudes
through various quark loop diagrams.

\subsection{Dirac structure of Hadron-quark vertex function for P-mesons in
BSE with power counting scheme}

\label{subsec:powercounting}

To obtain the form of Hadron-quark vertex function for the case of fermionic
quarks constituting a particular meson, we first replace the scalar
propagators $\Delta_{i}^{-1}$ in Eq. (\ref{eq:2.10}) by the proper fermionic
propagators $S_{F}$. The $H{q\bar{q}}$ vertex function $\Gamma(\hat{q})$ now
is a $4 \times4$ matrix in spinor space. For incorporation of the relevant
Dirac structures in $\Gamma(\hat{q})$, we make use of the power counting rule
we developed in \cite{bhatnagar06}, order-by-order in powers of inverse of
meson mass \cite{bhatnagar06}. Our aim of developing the power counting rule
was to find a ``criterion'' so as to systematically choose among various Dirac
covariants from their complete set to write wave functions for different
mesons (vector mesons, pseudoscalar mesons etc.).

As far as a pseudoscalar meson is concerned, its hadron-quark vertex function
which has a certain dimensionality of mass can be expressed as a linear
combination of four Dirac covariants \cite{smith69}, each multiplying a
Lorentz scalar amplitude, as function of $q\cdot P$.
%
We note that in the expression for CIA vertex function in equation
(\ref{eq:2.10}), the factor $D(\hat{q})\phi(\hat{q})$ is nothing but the
Lorentz-invariant momentum dependent scalar which depends on $q^{2}$, $P^{2}$
and $q\cdot P$ and has a certain dimensionality of mass. However the
Lorentz-scalar amplitudes multiplying various Dirac structures in
\cite{cvetic04} have different dimensionalities of mass. For adapting this
decomposition to write the structure of $H{q\bar{q}}$ vertex function
$\Gamma(\hat{q})$ for a particular meson, we re-express this function by
making these scalar amplitudes dimensionless by weighing each covariant by an
appropriate power $M$, the meson mass. Thus each term in the expansion of
$\Gamma(\hat{q})$ is associated with a certain power of $M$ and hence in
detail we can express the hadron-quark vertex, $\Gamma(\hat{q})$ as a
polynomial in various powers of $1/M$:
\begin{equation}
\Gamma^{P}(\hat{q})=\Omega^{P}\frac{1}{2\pi i}N_{P}D(\hat{q})\phi(\hat{q}),
\label{eq:2.12}%
\end{equation}
with
\begin{equation}
\Omega^{P}=\gamma_{5}B_{0}-i\gamma_{5}(\gamma\cdot P)\frac{B_{1}}{M}%
-i\gamma_{5}(\gamma\cdot q)\frac{B_{2}}{M}-\gamma_{5}[(\gamma\cdot
P)(\gamma\cdot q)-(\gamma\cdot q)(\gamma\cdot P)]\frac{B_{3}}{M^{2}},
\label{eq:2.121}%
\end{equation}
where $B_{i}$ $(i=0,...3)$ are four dimensionless coefficients
%
to be determined. Since we use constituent quark masses, where quark mass $m$
is approximately half of the hadron mass $M$, we can use the ansatz
\begin{equation}
q<<P\sim M \label{eq:2.13}%
\end{equation}
in the rest frame of the hadron (however we wish to mention that among all the
pseudoscalar mesons, pion enjoys the special status in view of its unusually
small mass ($M<\Lambda_{QCD}$) and its case should be considered separately).
Then each of the four terms in Eq. (\ref{eq:2.12},\ref{eq:2.121}) would again
receive suppression by different powers of $1/M$. Thus we can arrange these
terms as an expansion in powers of $O(1/M)$. We can then see in the expansion
of $\Omega^{P}$, that the structures associated with the coefficients $B_{0}$,
$B_{1}$ have magnitudes $O(1/M^{0})$ and are of leading order, while those
with $B_{2}$, $B_{3}$ are $O(1/M^{1})$ and are next-to-leading-order. This
na\"{i}ve power counting rule suggests that the maximum contribution to the
calculation of any pseudoscalar meson observable should come from the Dirac
structures $\gamma_{5}$ and $i\gamma_{5}(\gamma\cdot P)/M$ associated with the
constant coefficients $B_{0}$ and $B_{1}$ respectively, followed by the other
two higher order covariants associated with coefficients $B_{2}$ and $B_{3}$.
In general, the coefficients $B_{i}$ of the Dirac structures could be
functions of $q\cdot P$, and hence can be written as a Taylor series in powers
of $q\cdot P$. However the coefficients used here are dimensionless on lines
of \cite{bhatnagar06}. So they are in fact function of $q\cdot P/M^{2}$. Then
the leading order contribution of the coefficients are the case when the
$B_{i}$'s are constant. In this paper, we assume the coefficients are smooth
functions of $q\cdot P/M^{2}$, so to NLO, we only consider the terms of
eq.(10), with the coefficients $B_{i}$ constant. Because the normalization of
the BSW can be fixed (see below), $B_{0}$ here can be taken to be 1. So we
totally have 3 parameters to be fitted at NLO, comparing to one parameter at
LO. In a similar manner one can express the full hadron-quark vertex function
for a scalar and axial vector meson also in BSE under CIA. At the same time,
the restriction by charge parity on wave function of eigenstate should also be
respected. Further, to get the complete set of the Dirac structures for a
certain kind of meson, the restriction by the (space) Parity have been
employed; and it is easy to see that the requirements of the space Parity and
the charge Parity are the same for the vertex as well as the full wavefunction
\cite{bl07}. In this work to calculate the leptonic and radiative decay
constants, we take the form of hadron-quark vertex as in Eqs. (9) and (10)
which incorporates LO as well as NLO covariants and see the relative
importance of various covariants.

\subsection{BSE Kernel and the scalar wave function}

\label{subsec:BSEkernel}

From the above analysis of the structure of vertex function $H{q\bar{q}}$, we
notice that the structure of 3D wave function $\phi(\hat{q})$ as well as the
form of the 3D BSE are left untouched and have the same form as in our
previous works which justifies the usage of the same form of the input kernel
we used earlier \cite{bhatnagar06}. Now we briefly mention some features of
the BS formulation employed. The structure of BSE is characterized by a single
effective kernel arising out of a four-fermion lagrangian in the
Nambu-Jonalasino \cite{nambu61,mitra01} sense. The formalism is fully
consistent with Nambu-Jona-Lasino \cite{nambu61} picture of chiral symmetry
breaking but is additionally Lorentz-invariant because of the unique
properties of the quantity $\hat{q}^{2}$, which is positive definite
throughout the entire 4D space. The input kernel $K(q,q^{\prime})$ in BSE is
taken as one-gluon-exchange like as regards color [$(\bm{\lambda}%
^{(1)}/2)\cdot(\bm{\lambda}^{(2)}/2)$] and spin ($\gamma_{\mu}^{(1)}%
\gamma_{\mu}^{(2)}$) dependence. The scalar function $V(q-q^{\prime})$ is a
sum of one-gluon exchange $V_{OGE}$ and a confining term $V_{conf}$. Thus we
can write the interaction kernel as \cite{bhatnagar06,mitra01}:
\[
K(q,q^{\prime})=\left(  \frac{1}{2}\bm{\lambda}^{(1)}\right)  \cdot\left(
\frac{1}{2}\bm{\lambda}^{(2)}\right)  V_{\mu}^{(1)}V_{\mu}^{(2)}V(q-q^{\prime
});
\]%
\[
V_{\mu}^{(1,2)}=\pm2m_{1,2}\gamma_{\mu}^{(1,2)};
\]%

\begin{equation}%
\begin{array}
[c]{rcl}%
V(\hat{q}-\hat{q}^{\prime}) & = & \displaystyle\frac{4\pi\alpha_{S}(Q^{2}%
)}{(\hat{q}-\hat{q}^{\prime})^{2}}+\frac{3}{4}\omega_{q\bar{q}}^{2}\int
d^{3}\bm{r}\left[  r^{2}(1+4a_{0}\hat{m}_{1}\hat{m}_{2}M^{2}r^{2}%
)^{-1/2}-\frac{C_{0}}{\omega_{0}^{2}}\right]  e^{i(\bm{\hat{q}-\hat{q}'}%
)\cdot\bm{r}};\\
&  & \\
\alpha_{S}(Q^{2}) & = & \displaystyle\frac{12\pi}{33-2f}\left(  \ln\frac
{M_{>}^{2}}{\Lambda^{2}}\right)  ^{-1};\quad M_{>}=Max(M,m_{1}+m_{2}).
\end{array}
\label{eq:2.14}%
\end{equation}
The Ansatz employed for the spring constant $\omega_{q\overline{q}}^{2}$ in
Eq. (\ref{eq:2.14}) is \cite{bhatnagar06,mitra01},
\begin{equation}
\omega_{q\overline{q}}^{2}=4\widehat{m}_{1}\widehat{m}_{2}M_{>}\omega_{0}%
^{2}\alpha_{S}(M_{>}^{2}), \label{eq:2.15}%
\end{equation}
where $\widehat{m}_{1}$, $\widehat{m}_{2}$ are the Wightman-Garding
definitions of masses of constituent quarks defined earlier. Here the
proportionality of $\omega_{q\bar{q}}^{2}$ on $\alpha_{S}(Q^{2})$ is needed to
provide a more direct QCD motivation to confinement. This assumption further
facilitates a flavour variation in $\omega_{q\bar{q}}^{2}$. And $\omega
_{0}^{2}$ in Eq. (\ref{eq:2.14}) and Eq. (\ref{eq:2.15}) is postulated as a
universal spring constant which is common to all flavours. Here in the
expression for $V(\hat{q}-\hat{q}^{\prime})$, as far as the integrand of the
confining term $V_{conf}$ is concerned, the constant term $C_{0}/\omega
_{0}^{2}$ is designed to take account of the correct zero point energies,
while $a_{0}$ term ($a_{0}\ll1$) simulates an effect of an almost linear
confinement for heavy quark sectors (large $m_{1}$, $m_{2}$), while retaining
the harmonic form for light quark sectors (small $m_{1}$, $m_{2}$)
\cite{mitra01} as is believed to be true for QCD. Hence the term
$r^{2}(1+4a_{0}\widehat{m}_{1}\widehat{m}_{2}M_{>}^{2}r^{2})^{-1/2}$ in the
above expression is responsible for effecting a smooth transition from
harmonic ($q\overline{q}$) to linear ($Q\overline{Q}$) confinement. The basic
input parameters in the kernel are just four i.e. $a_{0}=0.028$, $C_{0}=0.29$,
$\omega_{0}=0.158$ GeV and QCD length scale $\Lambda=0.20$ GeV and quark
masses, $m_{u,d}=0.265$ GeV, $m_{s}=0.415$ GeV, $m_{c}=1.530$ GeV and
$m_{b}=4.900$ GeV which have been earlier fit to the mass spectrum of
$q\overline{q}$ mesons\cite{mitra01} obtained by solving the 3D BSE under
Null-Plane Ansatz (NPA). However due to the fact that the 3D BSE under CIA has
a structure which is formally equivalent to the 3D BSE under NPA, near the
surface $P.q=0$, the $q\overline{q}$ mass spectral results in CIA formalism
are exactly the same as the corresponding results under NPA
formalism\cite{mitra01,bhatnagar05}. The details of BS model under CIA in
respect of spectroscopy are thus directly taken over from NPA formalism (see
\cite{mitra01,bhatnagar05,bhatnagar06}. Now comes to the problem of the 3D BS
wave function. The ground state wave function $\phi(\hat{q})$ satisfies the 3D
BSE on the surface $P\cdot q=0$, which is appropriate for making contact with
O(3)-like mass spectrum (see \cite{mitra01}). Its fuller structure is
reducible to that of a 3D harmonic oscillator with coefficients dependent on
the hadron mass $M$ and the total quantum number $N$. The ground state wave
function $\phi(\hat{q})$ deducible from this equation thus has a gaussian
structure \cite{mitra01,bhatnagar06} and is expressible as:
\begin{equation}
\phi(\hat{q})\sim e^{-\hat{q}^{2}/2\beta^{2}}. \label{eq:2.16}%
\end{equation}
In the structure of $\phi(\hat{q})$ in (\ref{eq:2.16}), the parameter $\beta$
is the inverse range parameter which incorporates the content of BS dynamics
and is dependent on the input kernel $K(q,q^{\prime})$. \bigskip The structure
of the parameter $\beta$ in $\phi(\widehat{q})$ is taken as
\cite{mitra01,bhatnagar05,bhatnagar06}:
\begin{equation}
\beta^{2}=(2\hat{m}_{1}\hat{m}_{2}M\omega_{q\bar{q}}^{2}/\gamma^{2}%
)^{1/2};\gamma^{2}=1-\frac{2\omega_{q\bar{q}}^{2}C_{0}}{M_{>}\omega_{0}^{2}}.
\label{eq25}%
\end{equation}

We now give the calculation of leptonic decays constants of pseudoscalar
mesons employing both LO and NLO Dirac covariants according to our power
counting scheme in the framework discussed in next section.

%
%

\section{Calculations and Results for $f_{P}$}

\label{sec:3}

\subsection{Leptonic decays of pseudoscalar mesons to NLO}

\label{subsec:leptonicdecays}

Decay constants $f_{P}$ can be evaluated through the loop diagram which gives
the coupling of the two-quark loop to the axial vector current and can be
evaluated as:
\begin{equation}
f_{P}P_{\mu}=\langle0\vert\bar{Q}i\gamma_{\mu}\gamma_{5}Q\vert P(P)\rangle,
\label{eq:3.1}%
\end{equation}
which can in turn be expressed as a loop integral,
\begin{equation}
f_{P}P_{\mu}=\sqrt{3}\int d^{4}q\;Tr[\Psi_{P}(P,q)i\gamma_{\mu}\gamma_{5}].
\label{eq:3.2}%
\end{equation}
Bethe-Salpeter wave function $\Psi(P,q)$ for a P-meson is expressed as,
\begin{equation}
\Psi(P,q)=S_{F}(p_{1})\Gamma(\hat{q})S_{F}(-p_{2}), \label{eq:3.3}%
\end{equation}
which is expressed as the quark and anti-quark propagators flanking the
Hadron-quark vertex $\Gamma(\hat{q})$ function which is in turn expressed by
Eq. (\ref{eq:2.12},\ref{eq:2.121}).

Using $\Psi(P,q$) from Eq. (\ref{eq:3.3}), and incorporating $H{q\bar{q}}$
vertex function $\Gamma(\hat{q})$ from Eq. (\ref{eq:2.12},\ref{eq:2.121}) in
Eq. (\ref{eq:3.2}), evaluating trace over the gamma matrices and multiplying
both sides of Eq. (\ref{eq:3.2}) by $P_{\mu}/(-M^{2})$), we can express the
leptonic decay constant $f_{P}$ as,
\begin{equation}
f_{P}=f_{P}^{(0)}+f_{P}^{(1)}+f_{P}^{(2)}+f_{P}^{(3)}, \label{eq:3.5}%
\end{equation}
where $f_{P}^{(0)}$, $f_{P}^{(1)}$, $f_{P}^{(2)}$, $f_{P}^{(3)}$, are the
contributions to $f_{P}$ from the four Dirac covariants associated with
coefficients $B_{i}$ ($i=0,1,2,3$), and are expressed as:
\begin{equation}%
\begin{array}
[c]{rcl}%
f_{P}^{(0)} & = & \sqrt{3}N_{P}B_{0}\int d^{3}\bm{\hat{q}}D(\hat{q})\phi
(\hat{q})\int\limits_{-\infty}^{\infty}\displaystyle\frac{Md\sigma}{2\pi
i\Delta_{1}\Delta_{2}}\left[  -2m_{1}+2\frac{m_{1}^{3}}{M^{2}}-2m_{2}%
-2\frac{m_{1}^{2}m_{2}}{M^{2}}-2\frac{m_{1}m_{2}^{2}}{M^{2}}\right. \\
&  & \displaystyle\left.  +2\frac{m_{2}^{3}}{M^{2}}+4(m_{1}-m_{2}%
)\sigma\right]  ,\\
&  & \\
f_{P}^{(1)} & = & \sqrt{3}N_{P}B_{1}\int d^{3}\bm{\hat{q}}D(\hat{q})\phi
(\hat{q})\int\limits_{-\infty}^{\infty}\displaystyle\frac{Md\sigma}{2\pi
i\Delta_{1}\Delta_{2}}\left[  M-\frac{m_{1}^{4}}{M^{3}}+4\frac{m_{1}m_{2}}%
{M}+2\frac{m_{1}^{2}m_{2}^{2}}{M^{3}}-\frac{m_{2}^{4}}{M^{3}}-4\frac{\hat
{q}^{2}}{M}\right. \\
&  & \displaystyle\left.  +(m_{2}^{2}-m_{1}^{2})\sigma\frac{4}{M}-4M\sigma
^{2}\right]  ,\\
&  & \\
f_{P}^{(2)} & = & \sqrt{3}N_{P}B_{2}\int d^{3}\bm{\hat{q}}D(\hat{q})\phi
(\hat{q})\int\limits_{-\infty}^{\infty}\displaystyle\frac{Md\sigma}{2\pi
i\Delta_{1}\Delta_{2}}\left[  \frac{4}{M^{3}}(m_{1}^{2}-m_{2}^{2})\hat{q}%
^{2}\right. \\
&  & \displaystyle\left.  +\left(  M-\frac{m_{1}^{4}}{M^{3}}+4\frac{m_{1}%
m_{2}}{M}+2\frac{m_{1}^{2}m_{2}^{2}}{M^{3}}-\frac{m_{2}^{4}}{M^{3}}\right)
\sigma+4\frac{\hat{q}^{2}}{M}\sigma^{2}+\frac{4}{M}(m_{2}^{2}-m_{1}^{2}%
)\sigma^{2}-4M\sigma^{3}\right]  ,\\
&  & \\
f_{P}^{(3)} & = & \sqrt{3}N_{P}B_{3}\int d^{3}\bm{\hat{q}}D(\hat{q})\phi
(\hat{q})\int\limits_{-\infty}^{\infty}\displaystyle\frac{Md\sigma}{2\pi
i\Delta_{1}\Delta_{2}}\left(  -8\frac{m_{1}+m_{2}}{M^{2}}\hat{q}^{2}\right)
.\label{eq:3.6}%
\end{array}
\end{equation}
In deriving the above expressions, we had made use of the scalar products of
various momenta expressed in terms of integration variables $\hat{q}$ and
$\sigma$ as,
\begin{equation}%
\begin{array}
[c]{rcl}%
p_{1}\cdot p_{2} & = & -M^{2}(\hat{m_{1}}+\sigma)(\hat{m_{2}}-\sigma)-\hat
{q}^{3},\\
p_{1}\cdot P & = & -M^{2}(\hat{m_{1}}+\sigma),\\
p_{2}\cdot P & = & -M^{2}(\hat{m_{2}}-\sigma),\\
P\cdot q & = & -M^{2}\sigma,\\
p_{1}^{2} & = & -M^{2}(\hat{m_{1}}+\sigma)^{2}+\hat{q}^{2},\\
p_{2}^{2} & = & -M^{2}(\hat{m_{2}}-\sigma)^{2}+\hat{q}^{2},\\
p_{1}\cdot q & = & \displaystyle\frac{1}{2}\{2\hat{q}^{2}-\sigma\lbrack
m_{1}^{2}-m_{2}^{2}+M^{2}(1+2\sigma)]\},\\
p_{2}\cdot q & = & \displaystyle\frac{1}{2}\{-2\hat{q}^{2}+\sigma\lbrack
m_{1}^{2}-m_{2}^{2}+M^{2}(-1+2\sigma)]\}.
\end{array}
\label{eq:3.7}%
\end{equation}
We see that on the right hand side of the expression for $f_{P}$, each of the
expressions multiplying the constant parameters $B_{0}$ and $B_{1}$ consist of
two parts, of which only the second part explicitly involves the off-shell
parameter $\sigma$. It is can be seen that the off-shell contribution which
vanishes for $m_{1}=m_{2}$ in case of using only the leading covariant
$\gamma_{5}$, would no longer vanish for $m_{1}=m_{2}$ in the above
calculation for $f_{P}$ (when other covariants are incorporated in $H{q\bar
{q}}$ vertex function besides the leading covariant $\gamma_{5}$) due to the
terms like $4M$ and $4\hat{q}^{2}/M$ multiplying $\sigma^{2}$ in $f_{P}^{(2)}$
and $f_{P}^{(3)}$ respectively. This possibly implies that when other
covariants besides $\gamma_{5}$ are incorporated into the vertex function, the
off-shell part of $f_{P}$ does not arise from unequal mass kinematics alone
(which is in complete contrast to the earlier CIA calculation of $f_{P}$
employing only $\gamma_{5}$). This may be a pointer to the fact that Dirac
covariants other than $\gamma_{5}$ might also be important for the study of
processes involving large $q^{2}$ (off-shell). Carrying out integration over
$d\sigma$ by method of contour integration by noting the pole positions in the
complex $\sigma$-plane:
\begin{equation}%
\begin{array}
[c]{l}%
\displaystyle\Delta_{1}=0\Rightarrow\sigma_{1}^{\pm}=\pm\frac{\omega_{1}}%
{M}-\hat{m}_{1}\mp i\varepsilon,\quad\omega_{1}^{2}=m_{1}^{2}+\hat{q}^{2},\\
\\
\displaystyle\Delta_{2}=0\Rightarrow\sigma_{2}^{\mp}=\mp\frac{\omega_{2}}%
{M}+\hat{m}_{2}\pm i\varepsilon,\quad\omega_{2}^{2}=m_{2}^{2}+\hat{q}^{2},
\end{array}
\label{eq:3.8}%
\end{equation}
we can again express $f_{P}$ as $f_{P}=f_{P}^{(0)}+f_{P}^{(1)}+f_{P}%
^{(2)}+f_{P}^{(3)}$, where now
\begin{equation}%
\begin{array}
[c]{rcl}%
f_{P}^{(0)} & = & \displaystyle\sqrt{3}N_{P}B_{0}\int d^{3}\bm{\hat{q}}%
D(\hat{q})\phi(\hat{q})\left[  \left(  -2m_{1}+2\frac{m_{1}^{3}}{M^{2}}%
-2m_{2}-2\frac{m_{1}^{2}m_{2}}{M^{2}}-2\frac{m_{1}m_{2}^{2}}{M^{2}}\right.
\right. \\
&  & \\
&  & \displaystyle\left.  \left.  +2\frac{m_{2}^{3}}{M^{2}}\right)  \frac
{1}{D(\hat{q})}+4(m_{1}-m_{2})R_{1}\right]  ,\\
&  & \\
f_{P}^{(1)} & = & \displaystyle\sqrt{3}N_{P}B_{1}\int d^{3}\bm{\hat{q}}%
D(\hat{q})\phi(\hat{q})\left[  \left(  M-\frac{m_{1}^{4}}{M^{3}}+4\frac
{m_{1}m_{2}}{M}+2\frac{m_{1}^{2}m_{2}^{2}}{M^{3}}-\frac{m_{2}^{4}}{M^{3}%
}\right)  \frac{1}{D(\hat{q})}\right. \\
&  & \\
&  & \displaystyle\left.  -4\frac{\hat{q}^{2}}{M}\frac{1}{D(\hat{q})}+\frac
{4}{M}(m_{2}^{2}-m_{1}^{2})R_{1}-4MR_{2}\right]  ,\\
&  & \\
f_{P}^{(2)} & = & \displaystyle\sqrt{3}N_{P}B_{2}\int d^{3}\bm{\hat{q}}%
D(\hat{q})\phi(\hat{q})\left[  \frac{4}{M^{3}}(m_{1}^{2}-m_{2}^{2})\hat{q}%
^{2}\frac{1}{D(\hat{q})}\right. \\
&  & \\
&  & \displaystyle\left.  +\left(  M-\frac{m_{1}^{4}}{M^{3}}+4\frac{m_{1}%
m_{2}}{M}+2\frac{m_{1}^{2}m_{2}^{2}}{M^{3}}-\frac{m_{2}^{4}}{M^{3}}\right)
R_{1}\right. \\
&  & \\
&  & \displaystyle\left.  +4\frac{\hat{q}^{2}}{M}R_{1}+\frac{4}{M}(m_{2}%
^{2}-m_{1}^{2})R_{2}\right]  ,\\
&  & \\
f_{P}^{(3)} & = & \displaystyle\sqrt{3}N_{P}B_{3}\int d^{3}\bm{\hat{q}}%
D(\hat{q})\phi(\hat{q})\left[  -8\frac{1}{M^{2}}(m_{1}+m_{2})\hat{q}^{2}%
\frac{1}{D(\hat{q})}\right]  ,
\end{array}
\label{eq:3.9}%
\end{equation}
and $D(\hat{q})$ is given in Eq. (\ref{eq:2.9}), and the results of $\sigma
$-integration in the complex $\sigma$-plane, on whether the contour is closed
from above or below the real $\sigma$-axis is:
\begin{equation}%
\begin{array}
[c]{rcl}%
R_{1} & = & \displaystyle\int\limits_{-\infty}^{+\infty}\frac{Md\sigma}{2\pi
i\Delta_{1}\Delta_{2}}\sigma=\frac{M^{2}(-\omega_{1}+\omega_{2})+(m_{1}%
^{2}-m_{2}^{2})(\omega_{1}+\omega_{2})}{4M^{2}\omega_{1}\omega_{2}%
[M^{2}-(\omega_{1}+\omega_{2})^{2}]},\\
&  & \\
R_{2} & = & \displaystyle\int\limits_{-\infty}^{+\infty}{\frac{Md\sigma}{2\pi
i\Delta_{1}\Delta_{2}}\sigma^{2}}\\
& = & \displaystyle\frac{(-M^{4}-m_{12}^{2}\delta m^{2}+4M^{2}\omega_{1}%
\omega_{2})(\omega_{1}+\omega_{2})+2M^{2}m_{12}\delta m(\omega_{2}-\omega
_{1})}{8M^{4}\omega_{1}\omega_{2}[M^{2}-(\omega_{1}+\omega_{2})^{2}%
]}.\label{eq:3.10}%
\end{array}
\end{equation}
To calculate BS normalizer $N_{P}$ for a pseudoscalar meson in the expression
for $f_{P}$ in Eq. (\ref{eq:3.9}), we use the current conservation condition
\cite{bhatnagar06},
\begin{equation}
2iP_{\mu}=(2\pi)^{4}\int d^{4}q\;Tr\left[  \overline{\Psi}(P,q)\left(
\frac{\partial}{\partial P_{\mu}}S_{F}^{-1}(p_{1})\right)  \Psi(P,q)S_{F}%
^{-1}(-p_{2})\right]  +(1\Leftrightarrow2). \label{eq:3.11}%
\end{equation}
Putting BS wave function $\Psi(P,q)$ from Eq. (\ref{eq:3.3}) in the above
equation, carrying out derivatives of inverse of propagators of constituent
quarks with respect to total momentum of hadron $P_{\mu}$, evaluating trace
over the gamma matrices, following usual steps and multiplying both sides of
equation by $P_{\mu}/(-M^{2})$ to extract out the normalizer $N_{P}$ from the
above expression, we then express the above expression in terms of integration
variables $\hat{q}$ and $\sigma$. Noting that the four dimensional volume
element $d^{4}q=d^{3}\widehat{q}Md\sigma$, we then perform pole integration
over $d\sigma$ in complex $\sigma$-plane, making use of the pole positions in
Eq. (\ref{eq:3.8}). The calculation of normalizer is extremely complex due to
unequal mass kinematics. We thus give here a general expression for the
normalizer integral of the form,
\begin{equation}
N_{P}^{-1}=-(2\pi)^{2}i\int d^{3}\bm{\hat{q}}D^{2}(\hat{q})\phi^{2}(\hat
{q})[g_{1}(B,\hat{q})I_{1}+g_{2}(B,\hat{q})I_{2}+g_{3}(B,\hat{q})I_{3}%
+g_{4}(B,\hat{q})I_{4}], \label{eq:3.12}%
\end{equation}
where $B\equiv(B_{0},B_{1},B_{2},B_{3})$ and $g_{1},...g_{4}$ are extremely
complicated functions of $B$ and $\hat{q}$ and are extremely lengthy
expressions, and hence we do not present their actual forms here, whereas
$I_{1},...I_{4}$ are analytic results of pole integration over the off-shell
variable $\sigma$ in the complex $\sigma$-plane and are expressed as:

\[%
\begin{array}
[c]{rcl}%
I_{1} & = & \displaystyle\int\limits_{-\infty}^{+\infty}\frac{Md\sigma}%
{\Delta_{1}^{2}\Delta_{2}}=2\pi i\left[  \frac{2\omega_{1}^{3}-M^{2}\omega
_{2}+5\omega_{1}^{2}\omega_{2}+4\omega_{1}\omega_{2}^{2}+\omega_{2}^{3}%
}{4\omega_{1}^{3}\omega_{2}(M^{2}-(\omega_{1}+\omega_{2})^{2})^{2}}\right]
,\\
&  & \\
I_{2} & = & \displaystyle\int\limits_{-\infty}^{+\infty}\frac{Md\sigma}%
{\Delta_{1}^{2}\Delta_{2}}\sigma\\
&  & \\
& = & \displaystyle2\pi i\frac{-M^{4}\omega_{2}+(m_{1}^{2}-m_{2}^{2}%
)(\omega_{1}+\omega_{2})^{2}(2\omega_{1}+\omega_{2})M^{2}[6\omega_{1}%
^{3}+9\omega_{1}^{2}\omega_{2}+4\omega_{1}\omega_{2}^{2}+\omega_{2}(-m_{1}%
^{2}+m_{2}^{2}+\omega_{2}^{2})]}{8M^{2}\omega_{1}^{3}\omega_{2}[M^{2}%
-(\omega_{1}+\omega_{2})^{2}]^{2}},
\end{array}
\]%

\begin{equation}%
\begin{array}
[c]{rcl}%
I_{3} & = & \displaystyle\int\limits_{-\infty}^{+\infty}\frac{Md\sigma}%
{\Delta_{1}^{2}\Delta_{2}}\sigma^{2}\\
&  & \\
& = & \displaystyle2\pi i\frac{1}{16M^{4}\omega_{1}^{3}\omega_{2}%
(-M^{2}+(\omega_{1}+\omega_{2})^{2})^{2}}\{-M^{6}\omega_{2}+(m_{1}^{2}%
-m_{2}^{2})^{2}(\omega_{1}+\omega_{2})^{2}(2\omega_{1}+\omega_{2})\\
&  & \\
&  & \displaystyle+M^{4}[2\omega_{1}^{3}-2m_{1}^{2}\omega_{2}+2m_{2}^{2}%
\omega_{2}+\omega_{1}^{2}\omega_{2}+4\omega_{1}\omega_{2}^{2}+\omega_{2}%
^{3}]\\
&  & \\
&  & \displaystyle-M^{2}[m_{1}^{4}\omega_{2}+m_{2}^{4}\omega_{2}+4\omega
_{1}^{2}\omega_{2}(\omega_{1}+\omega_{2})^{2}+2m_{2}^{2}(-2\omega_{1}%
^{3}+\omega_{1}^{2}\omega_{2}+4\omega_{1}\omega_{2}^{2}+\omega_{2}^{3})\\
&  & \\
&  & \displaystyle-2m_{1}^{2}(-2\omega_{1}^{3}+m_{2}^{2}\omega_{2}+\omega
_{1}^{2}\omega_{2}+4\omega_{1}\omega_{2}^{2}+\omega_{2}^{3})]\}\\
&  & \\
I_{4} & = & \displaystyle\int\limits_{-\infty}^{+\infty}\frac{Md\sigma}%
{\Delta_{1}^{2}\Delta_{2}}\sigma^{3}\\
&  & \\
& = & \displaystyle2\pi i\left\{  \frac{(M^{2}-m_{1}^{2}+m_{2}^{2}%
+2M\omega_{2})^{3}}{8M^{6}\omega_{2}(M^{2}-\omega_{1}^{2}+2M\omega_{2}%
+\omega_{2}^{2})^{2}}\right. \\
&  & \\
&  & \displaystyle\left.  +\frac{(M^{2}+m_{1}^{2}-m_{2}^{2}-2M\omega_{1}%
)^{2}[M^{4}+M^{2}(m_{1}^{2}-m_{2}^{2}-\omega_{1}^{2}-\omega_{2}^{2}%
)+(m_{1}^{2}-m_{2}^{2})(3\omega_{1}^{2}-\omega_{2}^{2})]}{16M^{6}\omega
_{1}^{3}(M^{2}-2M\omega_{1}+\omega_{1}^{2}-\omega_{2}^{2})^{2}}\right\}  .\\
&  & \\
&  & \displaystyle\left.  +\frac{(M^{2}+m_{1}^{2}-m_{2}^{2}-2M\omega_{1}%
)^{2}[-4M\omega_{1}(m_{1}^{2}-m_{2}^{2}+\omega_{2}^{2})]}{16M^{6}\omega
_{1}^{3}(M^{2}-2M\omega_{1}+\omega_{1}^{2}-\omega_{2}^{2})^{2}}\right\}
.\label{eq:3.13}%
\end{array}
\end{equation}
After this, numerical integration over the 3-D variable $d^{3}\bm{\hat{q}}$ in
Eq. (\ref{eq:3.12}) is performed to evaluate $N_{P}$.

We have thus evaluated the expressions for $f_{P}$ and $N_{P}$ in framework of
BSE under CIA, with Dirac structures of eq. (\ref{eq:2.121}).
%
%
introduced in the $H{q\bar{q}}$ vertex function besides $\gamma_{5}$ according
to our power counting rule. We see that so far the results are independent of
any model for $\phi(\hat{q})$. However, for calculating the numerical values
of these decay constants one needs to know the constant coefficients $B_{0}$,
$B_{1}$, $B_{2}$, $B_{3}$ which are associated with the above Dirac
structures. Because of the normalization condition, we take $B_{0}=1$, and
then there are 3 parameters $B_{1}/B_{0}$, $B_{2}/B_{0}$, $B_{3}/B_{0}$, which
will still be denoted as $B_{1}$, $B_{2}$, $B_{3}$ for simplicity. To see the
contribution of various Dirac covariants on the calculation of meson decay
constants, we first discuss the numerical procedure adopted to fit these
coefficients
%
%
%

\subsection{Numerical Calculation}

\label{subsec:numerical}

Eq. (\ref{eq:3.9}) which expresses decay constants $f_{P}$ of pseudo-scalar
mesons in terms of the parameters $B_{0}$, $B_{1}$, $B_{2}$, $B_{3}$ is a
highly non linear function of the $B_{i}$'s. This obviously implies that
numerical methods must be applied to solve the problem.

We used a simple Mathematica procedure for calculating the numerical integrals
and searching for accurate values of the $B_{i}$ ($i=0,...3$). We defined the
following auxiliary function $W(B)$ which is positive definite as,
\begin{equation}
W(B)=\sum\limits_{P}[f_{P}(B)-f_{P}(exp.)]^{2}, \label{eq:3.14}%
\end{equation}
where $B\equiv(B_{0},B_{1},B_{2},B_{3})$, and summation in the above equation
runs over five pseudoscalar mesons $\pi$, $K$, $D$, $D_{S}$ and $B$ mesons
studied in this work, and $f_{P}(exp.)$ are the central values of experimental
data on decay constants \cite{eidelman04, babar} (indicated in Table II).

From the numerical point of view the problem reduces to finding values of
$B_{i}$'s such that $W(B)$ has a minimum. We used Mathematica package which
has some useful functions for minimizing. Those functions start from a point
and search for a minimum near to that initial point. We constrained all the
$Bi$'s to lie within the interval [0,1]. We generated in a random way values
of the $B_{i}$ in this interval. Starting from those values, the Mathematica
minimization function finds a minimum. Then it is checked if this minimum is
``sufficiently near to zero''. This check is done by evaluating the percent
average of the absolute values of the differences between the predicted
$f_{P}$ values from the experimental value $f_{P}(exp.)$. Using this method we
found that the values of coefficients $B_{0}$,...,$B_{3}$ (with average error
with respect to the experimental data less than 3.5\%) respectively are:
$B_{0}=1$, $B_{1}/B_{0}=0.3727$, $B_{2}/B_{0}=0.2234$, $B_{3}/B_{0}=0.0821$ to
give the decay constant values, $f_{\pi}=0.130$ GeV, $f_{K}=0.164$ GeV,
$f_{D}=0.194$ GeV, $f_{D_{s}}=0.296$ GeV. and $f_{B}=0.228$ GeV which are
within the error bars of experimental data \cite{eidelman04,babar} depicted in
Table II for these five pseudoscalar mesons. These values of $f_{P}$ along
with the contributions from various covariants and comparison with various
models and experimental results are listed in Tables I and II.

%
%
%

There is one important point which needs to be clarified: The experimental
data have different error bar, e.g., the data of $\pi$ has a very high
precision to the order of $0.1\%$, while for the case of $B$, the relative
error is more than $16\%$. So in the fitting, we should take into account the
difference, e.g., assign different weight for these data. However, we only
give our formulation at NLO. From the above discussions, it is straightforward
to rcognize, the smaller the meson mass, the larger the contributions of
higher orders. For the case of pion, we even can perspect that higher order
contributions (coming from higher order terms of Taylor series of
$B_{i}^{\prime}s$ as powers of $\frac{q.P}{M^{2}}$) could be also very
important. So, it is not reasonable to expect the NLO formulae can fit the
data of pion to the precision of order of $0.1\%$. This is the reason why we
fit the central value of data equally, as described above.

\section{Radiative Decay Constants of Neutral P--mesons}

\label{sec:4}

In this section we calculate the radiative decays of a neutral pseudoscalar
meson such as $\pi^{0}$ or $\eta_{c\text{ }}$proceeding through the process
$P\longrightarrow\gamma\gamma$ which proceed through the famous quark-triangle
diagrams in the above framework using both the leading order and the
next-to-leading order covariants in the Hadron-quark vertex function, taking
the values of parameters $B_{0}=1$, $B_{1}/B_{0}=0.3727$, $B_{2}/B_{0}%
=0.2234$, $B_{3}/B_{0}=0.0821$ fixed above in the calculation of $f_{P\text{
}}$values of $\pi,K,D,D_{S}$ and $B$ mesons. The invariant amplitude for the
decay of a neutral P-meson into two photons can be expressed as summation over
the two triangle diagrams corresponding to the Direct and Exchange processes as:%

\begin{equation}
A(P\rightarrow2\gamma)=\frac{e^{2}}{\sqrt{6}}\int d^{4}qTr[\overline{\Psi
}(P,q)i\gamma.\epsilon_{1}S_{F}(q-Q)i\gamma.\epsilon_{2}]+\frac{e^{2}}%
{\sqrt{6}}\int d^{4}qTr[\overline{\Psi}(P,q)i\gamma.\epsilon_{2}\bigskip
S_{F}(q+Q)i\gamma.\epsilon_{1} \label{3.15}%
\end{equation}
\ \ \ \ \ \ \ \ \ \ \ \ \ \ \ \ \ \ \ \ \ \ \ \ \ \ \ \ \ \ \ \ \ \ \ \ \ where
$\Psi(P,q)$ is the BS wave function of a neutral P-meson given explicitly in
Eq.(20) and Eq.(12)- (13), $S_{F}(q\pm Q)$ are the propagators of the third
quark in the Direct and Exchange diagrams respectively, where $Q=k_{1}-k_{2}$
is the the difference in momenta of the two emitted photons with momenta
$k_{1}$and $k_{2\text{ }}$respectively , while $\epsilon_{1,2}$ are the
polarization vectors of the two emitted photons in the above diagrams which
differ from each other in the interchange $1\Leftrightarrow2.$ Evaluating
traces over the gamma- matrices, combining various terms and then performing
pole-integrations in the complex $\sigma-$plane, we can express amplitude for
the above process as:

\bigskip%
\begin{equation}
A(P\rightarrow2\gamma)=[F_{P}]\epsilon_{\mu\nu\rho\sigma}P_{\mu}\epsilon
_{2\nu}Q_{\rho}\epsilon_{1\sigma}, \label{3.16}%
\end{equation}
\ \ \ \ \ \ \ \ \ \ \ \ \ \ \ \ \ \ \ \ \ \ \ \ \ \ \ \ \ \ \ \ \
\ \ \ \ \ \

where $P=p_{1}+p_{2}$ is the total hadron momentum, where $p_{1,2}$ are the
momenta of the quarks constituting the hadron, and the radiative decay
constant, $F_{P}$ is given as (the $B_{2}$ term vanishes because of wrong
charge parity),%

\begin{equation}
F_{P}=\frac{e^{2}N_{P}}{\sqrt{6}}\int d^{3}\widehat{q}D(\widehat{q}%
)\phi(\widehat{q})\left[  B_{0}[8mS_{1}]+B_{1}[\frac{-16m^{2}}{M}S_{1}%
+\frac{4}{M}S_{2}+\frac{4}{M}S_{3}]+B_{3}[\frac{8m}{M^{2}}(S_{2}+S_{3}%
+S_{4}-S_{5})]\right]  ,\ \label{3.17}%
\end{equation}
\ \ \ \ \ \ \ \ \ \ \ \ \ \ \ \ \ \ \ \ \ \ \ \ \ \ \ \ \ \ \ \ \ \ \ \ \ \ \ \ \ \ \ \ \ \ \ \ \ where
$S_{1,2,3,4,5}$ are the analytical results of integrals over the off-shell
parameter $\sigma$:

\bigskip%
\begin{align}
S_{1} &  =\int_{-\infty}^{+\infty}\frac{Md\sigma}{2\pi i\Delta_{1}\Delta
_{2}\Delta_{3}}=\frac{12}{M^{4}\omega-20M^{2}\omega^{3}+64\omega^{5}};\ \\
S_{2} &  =\int_{-\infty}^{+\infty}\frac{Md\sigma}{2\pi i\Delta_{2}\Delta_{3}%
}=\frac{4}{-M^{2}\omega+16\omega^{3}};\bigskip\\
S_{3} &  =\int_{-\infty}^{+\infty}\frac{Md\sigma}{2\pi i\Delta_{1}\Delta_{3}%
}=\frac{4}{-M^{2}\omega+16\omega^{3}};\nonumber\\
S_{4} &  =\int_{-\infty}^{+\infty}\frac{Md\sigma}{2\pi i\Delta_{1}\Delta_{3}%
}\sigma=\frac{-1}{-M^{2}\omega+16\omega^{3}}\\
S_{5} &  =\int_{-\infty}^{+\infty}\frac{Md\sigma}{2\pi i\Delta_{2}\Delta_{3}%
}\sigma=\frac{1}{-M^{2}\omega+16\omega^{3}}%
\end{align}

\bigskip

evaluated by the method of contour integrations by noting the various pole
positions in the complex $\sigma-$plane :

\bigskip%
\begin{align}
\Delta_{1}  &  =0\Rightarrow\sigma_{1}^{\pm}=\pm\frac{\omega}{M}-\frac{1}%
{2}\mp i\varepsilon;\\
\Delta_{2}  &  =0\Rightarrow\sigma_{2}^{\pm}=\pm\frac{\omega}{M}+\frac{1}%
{2}\mp i\varepsilon;\nonumber\\
\Delta_{3}  &  =0\Rightarrow\sigma_{3}^{\pm}=\pm\frac{\omega}{M}\mp
i\varepsilon;\omega^{2}=m^{2}+\widehat{q}^{2}\nonumber
\end{align}

corresponding to inverse propagators of the three quarks (of which
$\Delta_{1,2}$ correspond to the two constituent quarks in the meson) in the
quark-triangle diagrams, expressed in terms of the off-shell parameter
$\sigma$as:

\bigskip%
\begin{align}
\Delta_{1}  &  =\omega^{2}-M^{2}(\frac{1}{2}+\sigma);\\
\Delta_{2}  &  =\omega^{2}-M^{2}(\frac{1}{2}-\sigma);\nonumber\\
\Delta_{3}  &  =\omega^{2}-M^{2}\sigma^{2}\nonumber
\end{align}

\ \ \ \ \ \ \ \ \ \ \ \ \ \ \ \ \ \ \ \ \ \ \ \ \ \ \ \ \ \ \ \ \
\ \ \ \ \ \ \ \ \ \ \ \ \ \ \ \ \ \ \ \ \

\bigskip

From Eq.(33), it can be noticed that the contribution to radiative decay
constant $F_{P}$ from one of the next-to-leading order covariants associated
with the parameter $B_{2}$completely vanishes after trace evaluation.
Numerical evaluation of $F_{P}$ for $\pi^{0}$ and $\eta_{c}$ using the same
set of parameters, $B_{i\text{ }}/B_{0}$ fixed from the calculation of
leptonic decay constant $f_{P\text{ }}$values of $\pi,K,D,D_{S}$ and $B$
mesons above gives $F_{\pi}=.031GeV^{-1},F_{\eta_{c}}=.006GeV^{-1}$. These are
very close to the experimental numbers $F_{\pi}(Exp.)=.025GeV^{-1}$, and
$F_{\eta_{c}}(Exp.)=.0074GeV^{-1}$ which are arrived at through the
expression, $\Gamma(P\rightarrow2\gamma)=\frac{F_{P}^{2}M^{3}}{64\pi}$
connecting the decay width $\Gamma$ with radiative decay constants, $F_{P}$,
using the central values of experimental data on decay widths for $\pi$ and
$\eta_{c\text{ }}$mesons as $\Gamma(\pi^{0}\rightarrow2\gamma)=8.5eV$ and
$\Gamma(\eta_{c}\rightarrow2\gamma)=7.4KeV$ \cite{groom00,lansberg08} respectively.\bigskip

\section{Discussion}

\label{sec:5}

In this paper we have calculated the decay constants $f_{P}$ of pseudoscalar
mesons $\pi$, $K$, $D$, $D_{S}$ and $B$ \ and radiative decay constants
$F_{P}$ for neutral pseudoscalar mesons $\pi^{0}$ and $\eta_{c}$ proceeding
through the process $P\rightarrow2\gamma$ in BSE under CIA. The Hadron-quark
vertex function incorporates various Dirac covariants order-by-order in powers
of inverse of meson mass within its structure in accordance with a power
counting rule from their complete set. This power counting rule suggests that
the maximum contribution to any meson observable should come from Dirac
structures associated with Leading order terms alone, followed by Dirac
structures associated with Next-to-Leading Order terms in the vertex function.
Incorporation of all these covariants is found to bring calculated $f_{P}$
values much closer to results of experimental data \cite{eidelman04,babar} and
some recent calculations \cite{cvetic04,alkofer02,follana06,narison02} for
$\pi$, $K$, $D$, $D_{S}$ and $B$ mesons. The $f_{P}$ are within the error bars
of experimental data for each one of these five mesons by fitting three
parameters. The calculation of radiative decay constants of $\pi^{0}$ and
$\eta_{c}$ is again close to the experimental data\cite{groom00,lansberg08}.

The results for $\pi$, $K$, $D$, $D_{S}$ and $B$ mesons with parameter set:
$B_{0}=1$, $B_{1}/B_{0}=0.3727$, $B_{2}/B_{0}=0.2234$, $B_{3}/B_{0}=0.0821$
(giving $f_{P}$ values with average error with respect to experimental data
less than 3.5\%) are presented in Table I. In Fig. 1 we are plotting functions
$I_{P}^{i}(\hat{q})$ ($i=0,...3$) vs $\hat{q}$, where $I_{P}^{i}(\hat{q})$ is
the integrand of $f_{P}^{(i)}$ in equations (\ref{eq:3.9}). The plots of
variations of $I_{P}^{0}(\hat{q}),...I_{P}^{3}(\hat{q})$ with $\hat{q}$ for
$\pi$, $K$, $D$, $D_{S}$ and $B$ mesons, along with the results in Table I,
show that the contribution to $f_{P}$ from NLO covariants is much smaller than
the contribution from LO covariants for $K$, $D$, $D_{S}$ and $B$ mesons.
Comparison with experimental data and other models is shown in Table II. It is
seen from Table I that as far as the various contributions to decay constants
$f_{P}$ are concerned, for $K$ mesons, the LO terms contribute 60\%, while NLO
terms 40\%. However for heavy-light meson $D$, the LO contribution increases
to 90\%, while NLO contribution is 10\%. For $D_{S}$ meson, LO contribution is
91\%, while NLO contribution is 9\%. But for $B$ meson, the LO contribution is
96\%, while NLO contribution reduces to just 4\%. This is in conformity with
the power counting rule according to which the leading order covariants,
$\gamma_{5}$ and $i\gamma_{5}(\gamma\cdot P)(1/M)$ (associated with
coefficients $B_{0}$ and $B_{1}$) should contribute maximum to decay constants
followed by the next-to-leading order covariants, $-i\gamma_{5}(\gamma\cdot
q)(1/M)$ and $-\gamma_{5}[(\gamma\cdot P)(\gamma\cdot q)-(\gamma\cdot
q)(\gamma\cdot P)](1/M^{2})$ (associated with coefficients $B_{2}$ and $B_{3}%
$) in the BS wave function, Eq. (\ref{eq:2.12})-(\ref{eq:2.121}).

However the situation is different for the lightest meson $\pi$ which \bigskip
enjoys a unique status due to the fact that mass of a pion, $M$ is unusually
small ($<<\Lambda_{QCD}$), and the large difference between the sum of two
constituent quark masses and the pion mass shows that the quarks are far off
shell and the internal momentum $q$ should be the same order as the pion mass
and the approximation $q<<P\sim M$ breaks down for pion. Hence the
contribution of NLO covariants in pion case is even larger than the
contribution of LO covariants. Thus, the NLO covariants in pion should play a
more dominant role in contrast to heavier mesons $K$, $D$, $D_{S}$ and $B$.
However the sum of LO and NLO contributions adds up to the experimental value
for pion $f_{P}$ (=0.130 GeV). Further investigations on higher order terms
can show even more details of the pion structure.

To check the validity of our calculation, we then do numerical evaluation of
radiative decay constants$\ F_{P}$ for $\pi^{0}$ and $\eta_{c}$ using the same
set of parameters, $B_{i}/B_{0}$fixed above from the calculation of leptonic
decay constant $f_{P\text{ }}$values of $\pi,K,D,D_{S}$ and $B$ mesons. This
gives $F_{\pi}=.031GeV^{-1},F_{\eta_{c}}=.006GeV^{-1}$. These are very close
to the experimental numbers $F_{\pi}(Exp.)=.025GeV^{-1}$, and $F_{\eta_{c}%
}(Exp.)=.0074GeV^{-1}$ which are arrived at through the expression,
$\Gamma(P\rightarrow2\gamma)=\frac{F_{P}^{2}M^{3}}{64\pi}$ connecting the
decay width $\Gamma$ with radiative decay constants, $F_{P}$, using the
central values of experimental data on decay widths for $\pi$ and
$\eta_{c\text{ }}$mesons as $\Gamma(\pi^{0}\rightarrow2\gamma)=8.5eV$ and
$\Gamma(\eta_{c}\rightarrow2\gamma)=7.4$ \cite{groom00,lansberg08}
respectively.\bigskip\

The numerical results for leptonic decay constants, $f_{P}$ and radiative
decay constants, $F_{P}$ obtained in our framework upto the next to leading
order covariants demonstrates the validity of our power counting rule, which
also provides a practical means of incorporating various Dirac covariants in
the BS wave function of a hadron. By this rule, we also get to understand the
relative importance of various covariants to calculate various meson
observables. This would in turn help in obtaining a better understanding of
the hadron structure. Here would would like mention the robustness of our
framework: On one hand, at lower order(s), with limited number of parameters,
we can globally reproduce almost all the decay constants of certain kinds of
meson. On the other hand, by introducing higher order corrections, we can
accommodate enough parameters to fit the data as precise as possible, so than
to get a good parameterization of the structure of certain special hadron for
further investigations.


\begin{acknowledgments}
Major part of this work was carried out at Abdus Salam ICTP during the
Associateship visit of SB and JM during July-September 2008. We thank ICTP for
hospitality and for facilities provided during the course of this work. SB
also thanks Addis Ababa University where a good part of work on radiative decays was done.
JM thanks the support from University of Antioquia. SYL is partially supported
by NSFC with grant nos. 10775090 and 10935012 and NSF of
Shandong Province, China with account no. 2009ZRB02398.
\end{acknowledgments}

%
%
%
%

\newpage

%
%
%
 \begin{table}[h]
\begin{center}%
\begin{tabular}
[c]{|c|c|c|c|c|c|c|c|c|c|}\hline
& $f_{P}^{0}$ & $f_{P}^{1}$ & $f_{P}^{2}$ & $f_{P}^{3}$ & $|f_{P}^{LO}|$ &
$|f_{P}^{NLO}|$ & $f_{P}^{LO}$(\%) & $f_{P}^{NLO}$(\%) & $\bm{f_P=f_P^{LO}%
+ f_P^{NLO}}$\\\hline
$\pi$ & 0.110 & -0.154 & 0.000 & 0.175 & 0.044 & 0.175 & 25\% & 75\% &
\textbf{0.130}\\\hline
$K$ & 0.202 & -0.104 & 0.025 & 0.039 & 0.098 & 0.064 & 60\% & 40\% &
\textbf{0.164}\\\hline
$D$ & 0.271 & -0.097 & 0.010 & 0.009 & 0.174 & 0.019 & 90\% & 10\% &
\textbf{0.194}\\\hline
$D_{S}$ & 0.426 & -0.156 & 0.013 & 0.013 & 0.270 & 0.026 & 91\% & 9\% &
\textbf{0.296}\\\hline
$B$ & 0.345 & -0.125 & 0.005 & 0.003 & 0.220 & 0.008 & 96\% & 4\% &
\textbf{0.228}\\\hline
\end{tabular}
\end{center}
\caption{{\small {Decay constant $f_{P}$ values (in GeV) for $\pi$ $,$K $,$D
$D_{S}$ and $B$ mesons in BSE with the individual contributions $f_{p}^{0}$,
$f_{p}^{1}$, $f_{p}^{2}$, $f_{p}^{3}$ from various Dirac covariants along with
the contributions from LO and NLO covariants and also their \% contributions
for parameter set: $B_{0} = 1$, $B_{1}/B_{0} = 0.3727$, $B_{2}/B_{0} =
0.2234$, $B_{3}/B_{0} = 0.0821$ (with average error with respect to the
experimental data less than 3.5\%)}}}%
\label{tab:LONLO35}%
\end{table}

\begin{table}[h]
\begin{center}%
\begin{tabular}
[c]{|c|c|c|c|c|c|}\hline
& $f_{\pi}$ & $f_{K}$ & $f_{D}$ & $f_{D_{S}}$ & $f_{B}$\\\hline
BSE (3.5\% average error) &  &  &  &  & \\
present paper & \textbf{0.130} & \textbf{0.164} & \textbf{0.194} &
\textbf{0.296} & \textbf{0.228}\\\hline
BSE \cite{cvetic04} &  &  &  & 0.248 & \\\hline
SDE \cite{alkofer02} &  & 0.164 &  &  & \\\hline
Lattice \cite{follana06} &  &  & 0.208$\pm$0.004 & 0.241$\pm$0.003 & \\\hline
QCD-SR \cite{narison02} &  &  & 0.20$\pm$0.02 & 0.23$\pm$0.02 & \\\hline
Exp. Results \cite{eidelman04} & 0.1300$\pm$0.0001 & 0.159$\pm$0.001 &
0.22$\pm$0.02 & 0.29$\pm$0.03 & \\\hline
Babar+Belle &  &  &  &  & \\
Collaboration \cite{babar} &  &  &  &  & 0.24$\pm$0.04\\\hline
\end{tabular}
\end{center}
\caption{{\small {Comparison of results of $f_{P}$ (in GeV) for $\pi$ $K$,
$D$, $D_{S}$ and $B$ in BSE with the parameter set $B_{0} = 0.7045$, $B_{1} =
0.2626$, $B_{2} = 0.1574$, $B_{3} = 0.0579$ (with average error 3.5\%) with
those of other models and experimental data.}}}%
\label{tab:comparison}%
\end{table}

%
%
%
%
%
%
%
%
%

%
%

%
%

%

%
%
%
%
%
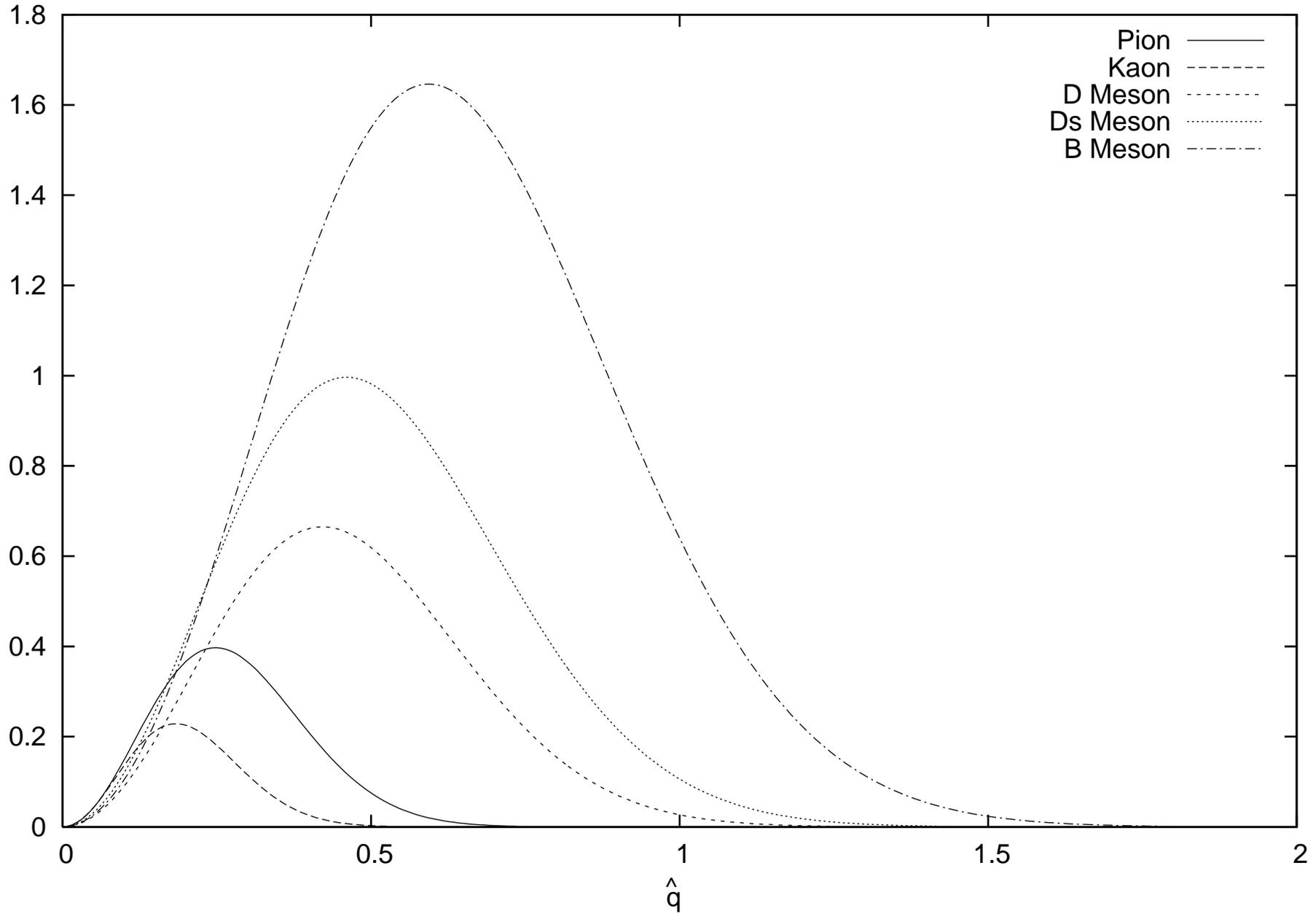
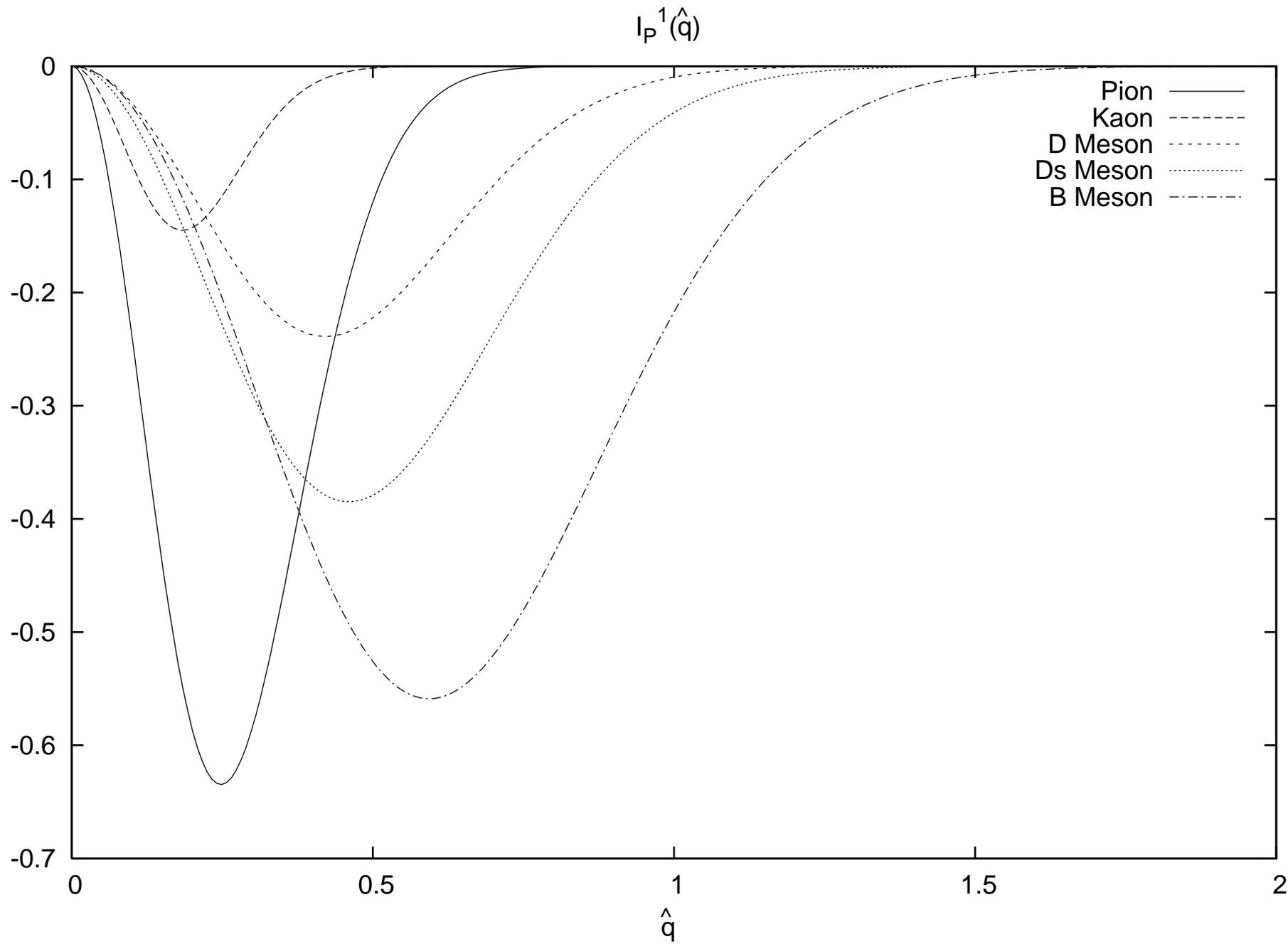
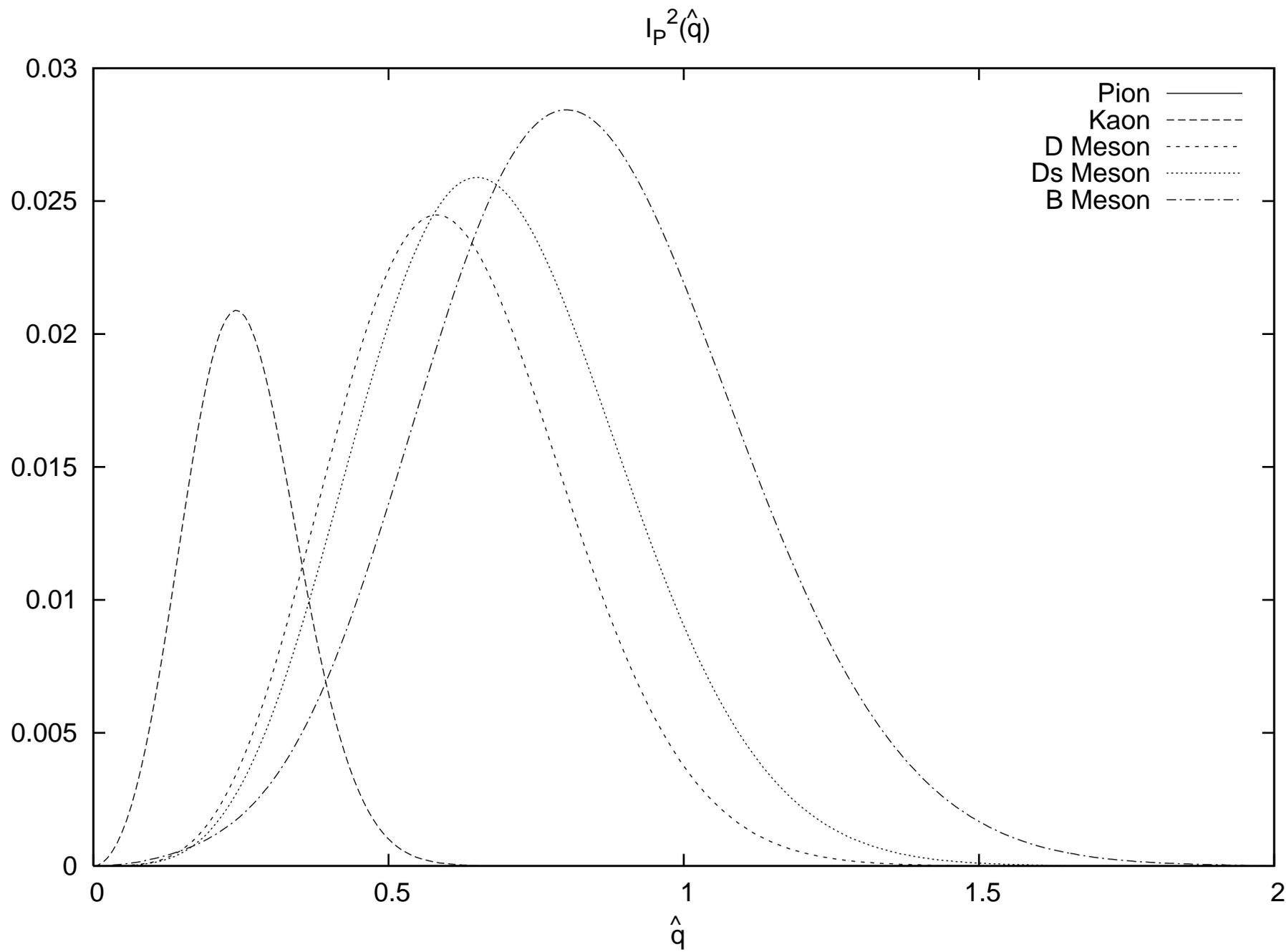
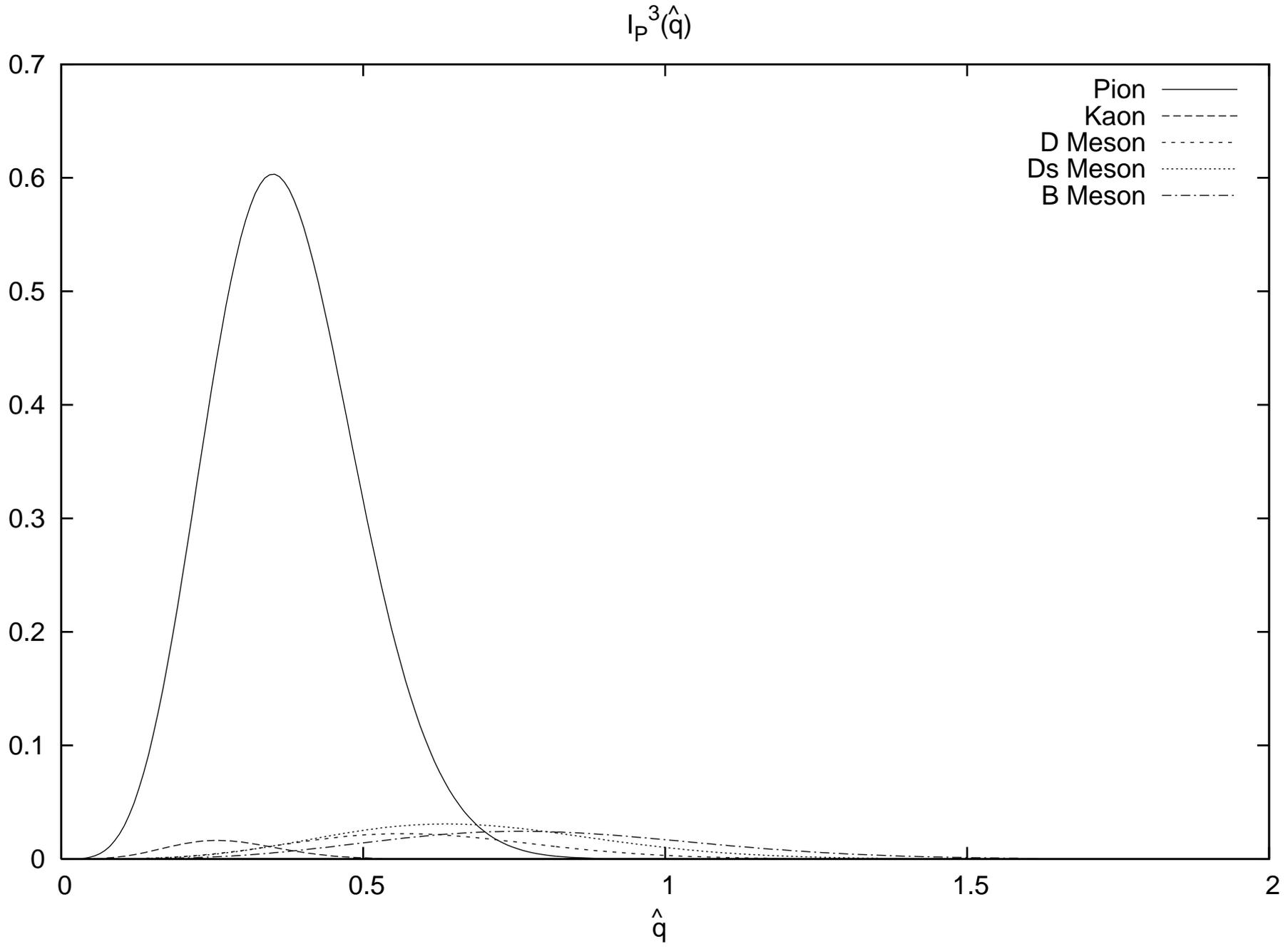
\end{document}